\documentclass[useAMS,usenatbib,usegraphicx]{mn2e}
\usepackage{subfigure}

\newcommand\vect[1]{{\mbox{\boldmath $#1$}}}
\newcommand\uVec{\vect{\rm u}}
\newcommand\chiVec{\vect{\chi}}
\newcommand\xVec{\vect{\rm x}}
\newcommand\AMat{\vect{\rm A}}
\newcommand\bVec{\vect{\rm b}}
\newcommand\gradPsi{\grad\psi}
\newcommand\grad{\vect{\nabla}}

\newcommand\kmax{k_{\rm max}}

\title[Multipole Models of Gravitational Lenses]
{Multipole Models of Four-Image Gravitational Lenses with Anomalous Flux Ratios}

\author[Arthur B. Congdon \& Charles R.\ Keeton]
{Arthur B. Congdon and Charles R.\ Keeton \\
Department of Physics and Astronomy, Rutgers University,
136 Frelinghuysen Rooad, Piscataway, NJ 08854 USA}

\date{Accepted in MNRAS}

\begin{document}

\maketitle

\begin{abstract}
It has been known for over a decade that many four-image
gravitational lenses exhibit anomalous radio flux ratios. These
anomalies can be explained by adding a clumpy cold dark matter (CDM)
component to the background galactic potential of the lens. As an
alternative, \citet{Evans_flux_ratios} recently suggested that smooth
multipole perturbations provide a reasonable alternative to CDM
substructure in some but not all cases.  We generalize their method
in two ways so as to determine whether multipole models can explain
highly anomalous systems.  We carry the multipole expansion to
higher order, and also include external tidal shear as a free
parameter. Fitting for the shear proves crucial to finding a
physical (positive-definite density) model.  For B1422+231, working
to order $k_{\rm max} = 5$ (and including shear) yields a model that
is physical but implausible. Going to higher order ($k_{\rm max} \ga
9$) reduces global departures from ellipticity, but at the cost of
introducing small scale wiggles in proximity to the bright images.
These localized undulations are more pronounced in B2045+265, where
$\kmax \sim 17$ multipoles are required to smooth out large scale
deviations from elliptical symmetry. Such modes surely cannot be
taken at face value; they must indicate that the models are trying
to reproduce some other sort of structure. Our formalism naturally
finds models that fit the data exactly, but we use B0712+472 to show
that measurement uncertainties have little effect on our results.
Finally, we consider the system B1933+503, where two sources are
lensed by the same foreground galaxy. The additional constraints
provided by the images of the second source render the multipole
model unphysical. We conclude that external shear
must be taken into account to obtain plausible models, and that a
purely smooth angular structure for the lens galaxy does not provide
a viable alternative to the prevailing CDM clump hypothesis.
\end{abstract}


\section{Introduction}

It has been suspected since the time of Newton that a light ray
would be deflected when it passes near a massive object. This
phenomenon, now known as gravitational lensing, provided one of the
earliest and best tests of the validity of general relativity. Since
that time, lens theory has matured into an active field of astronomy
\citep[see][for a recent review]{Kochanek_review}. Probing the
structure of galaxies is one of the many areas to which lensing has
been applied over the past fifteen years
\citep[e.g.,][]{Kochanek_galStruc, Keeton_structure}. Because lensing
is sensitive to all mass, it is possible to study both dark and
luminous components within galaxies.

In the case of strong lensing where the light source is multiply
imaged, we can use the positions and fluxes of the images to study
small-scale structure within the lens galaxy. This technique can be
most readily applied to four-image systems in a cusp or fold
configuration, which occurs when the angular separation between the
lens and source is small. For a cusp lens we expect the three
brightest images to satisfy the magnification sum rule
\citep[e.g.,][]{Schneider_cusp, Mao_substructure, Keeton_cusp},
\begin{equation}
\mu_1 + \mu_2 + \mu_3 \approx 0.
\end{equation}
In the fold case an analogous relation holds for the two brightest
images \citep{Gaudi_caustic, Keeton_fold}:
\begin{equation}
\mu_1 + \mu_2 \approx 0.
\end{equation}

Although these relations should hold for all smooth lens potentials,
there are a number of observed systems for which they are violated
\citep{Mao_substructure, Keeton_cusp, Keeton_fold}. Since the
magnifications are not directly observable, we refer to systems that
violate the sum rule as exhibiting flux ratio anomalies, with the
observed fluxes being related to the magnifications by the flux of
the source.

One might argue that violations of the magnification sum rule
originate in electromagnetic effects of the interstellar medium on
light emitted by the lensed source. However, dust extinction is
negligible at radio wavelengths, which are much larger than the
typical size of a dust grain.  The lack of wavelength dependence in
radio flux ratios rules out electromagnetic effects as an
explanation for radio anomalies \citep[see e.g.,][and references
therein]{Kochanek_multipole_ruled_out, Keeton_cusp, Keeton_fold}.

Explaining the observed anomalies therefore requires the addition of
small-scale structure to the gravitational potential of the galaxy
\citep{Mao_substructure}. A possible candidate for this substructure
emerged through the work of \citet{Moore_substructure} and
\citet{Klypin_missing}, whose numerical simulations predicted a
quantity of cold dark matter (CDM) halos with masses $\sim 10^6 -
10^8 M_{\odot}$, greatly exceeding the observed numbers of such
objects. This result motivated \citet{Metcalf_substructure} to
consider how the abundance of predicted CDM substructure might
affect lensing.  They pointed out that even if the small halos are
completely dark --- invisible to standard luminosity-based studies
--- they could still affect lens flux ratios and perhaps explain the
observed anomalies. Indeed, \citet{Dalal_substructure} found that
putting $\sim 2\%$ of the mass in $\sim 10^6 M_{\odot}$ halos could
reproduce the observed flux ratios for seven anomalous lens systems,
while broadly matching the predictions of the numerical simulations.
But does this mean that the flux ratio problem is really solved, and
that the ``missing" CDM satellites have been found? Are there other
plausible models that can solve the flux ratio problem?

Possibilities may include stellar microlensing, and complex
structure such as isophote twists or triaxiality in the lens galaxy.
Since the radio-emitting regions of the QSO sources we will consider
have a much larger angular scale than the Einstein radius of a
typical star in the lens galaxy, microlensing can be eliminated as a
potential cause of flux ratio anomalies. That leaves the question of
whether models that alter the global structure of the lens potential
offer a viable explanation of flux ratio anomalies. Our goal is to
see whether we can fit four anomalous radio lenses using models with
general but reasonable angular structure.

We begin with the self-similar multipole model of
\citet{Evans_flux_ratios}. In this framework, the potential of the
lens galaxy is described by a generalized isothermal model whose
angular dependence is expressed as a Fourier series. The multipole
coefficients are determined by fitting the observed image positions
and flux ratios. The truncation order of the series is chosen so
that the matrix of constraints is square. \citet{Evans_flux_ratios}
found that such a model could explain some but not all observed
lenses. In particular, for B1422+231 (the only radio anomaly they
studied), they could find a physically acceptable model only by
inflating the errorbars on the data, and even that model was highly
non-elliptical and implausible.

In this paper we extend the multipole formalism to include external
shear --- tidal distortions from objects near the lens galaxy
\citep[e.g.,][]{Keeton_shear} --- and higher order multipole modes.
Shear in particular will prove essential for obtaining sensible
solutions.  While this is not the most general method, it is
mathematically simple and may be a reasonable alternative to
substructure in some lenses.

As a test case, we first apply our model to Q2237+0305, which is not
anomalous at radio wavelengths. Then, we apply the model to
B1422+231, B2045+265, B0712+472, and B1933+503, which are all highly
anomalous. An exhaustive study of the known radio anomalies would
also include B1555+375, but the position of the lens galaxy (a key
ingredient in our formalism) is unknown.

\section{Methods}
\label{sec:methods}

We begin by writing the convergence (dimensionless surface mass
density), $\kappa$, and lens potential, $\psi$, for a galaxy with a
flat rotation curve and arbitrary angular structure. This model is
often referred to as a generalized isothermal model
\citep{Witt_isothermal, Zhao_isothermal, Evans_isothermal,
Evans_flux_ratios}:
\begin{equation}
    \kappa(r, \theta) = \frac{1}{2\,r}\,G(\theta)\,; \qquad \psi(r,
    \theta) = rF(\theta)\,. \label{eqn:kappa and psi}
\end{equation}
Noting that $\nabla^2 \psi = 2 \kappa$, we find that $F$ and $G$ are
related by
\begin{equation}
G(\theta) = F(\theta) + F''(\theta)\,. \label{eqn:general G}
\end{equation}
For a given source position, $\uVec \equiv (u,v)$, we can find the
image positions, $\xVec \equiv (r\cos{\theta},r\sin{\theta})$, via
the lens equation,
\begin{equation}
    \uVec = \xVec - \gradPsi(\xVec)\,.
    \label{eqn:lens equation}
\end{equation}
The inverse magnification of an image at $\xVec$ is given by
\begin{equation}
    \mu^{-1} ={\rm det} \left(\frac{\partial\uVec}{\partial\xVec}\right) = 1 - \frac{G(\theta)}{r}\,.
    \label{eqn:inverse magnification}
\end{equation}

An important property of the lens potential is the critical curve,
along which the magnification is formally infinite. The critical
curve in the image plane maps to the caustic in the source plane,
which marks the transition between 2-image and 4-image systems. We
see from equation (\ref{eqn:inverse magnification}) that the
critical curve is given by $r_{\rm crit}(\theta) = G(\theta)$, which
is equivalent to the isodensity contour with $\kappa = 1/2$.

\subsection{Multipole Lens Model of \citet{Evans_flux_ratios}}

Let us write the angular part of the potential, $F$, as a multipole
(Fourier) expansion in $\theta$, i.e.
\begin{equation}
    F (\theta) \equiv \frac{a_0}{2} + \sum_{k =1}^{k_{{\rm max}}}
    \left(a_k \cos k \theta + b_k \sin k \theta \right),
    \label{eqn:expanded F}
\end{equation}
for some appropriate $k_{\rm max}$. The Fourier coefficient $a_{0}$
is related to the Einstein radius by $R_{\rm Ein} \equiv a_{0}/2$.
We can find the unknown source position, $\uVec$, and the Fourier
coefficients, $a_k$ and $b_k$, by introducing observational
constraints, viz. the image positions and flux ratios.

From equations (\ref{eqn:kappa and psi}) and (\ref{eqn:expanded F}),
the lens equation (\ref{eqn:lens equation}) becomes
\begin{eqnarray}
    \label{eqn:u equation}
    u &=& r_l\,\cos\theta_l -\frac{a_0}{2}\alpha_0(\theta_l) -
    \sum_{k=1}^{k_{\rm max}}\left[a_k\alpha_k(\theta_l) +
    b_k\beta_k(\theta_l)\right]\,, \\
    \label{eqn:v equation}
    v &=& r_l\,\sin\theta_l -\frac{a_0}{2}\hat{\alpha}_0(\theta_l) -
    \sum_{k=1}^{k_{\rm max}}\left[a_k\hat{\alpha}_k(\theta_l) +
    b_k\hat{\beta}_k(\theta_l)\right]\,,
\end{eqnarray}
where $l = 1, 2, \, \ldots \,, n$ is the image number. The functions
$\alpha_k$, $\hat{\alpha}_k$, $\beta_k$, and $\hat{\beta}_k$ are
defined in \citet{Evans_flux_ratios}, equations (13) and (14):
\begin{eqnarray}
    \alpha_k(\theta) &=& \cos{\theta}\,\cos{k\theta}\,+\,k\,\sin{\theta}\,\sin{k\theta}\,, \\
    \hat{\alpha}_k(\theta) &=& \sin{\theta}\,\cos{k\theta}\,-\,k\,\cos{\theta}\,\sin{k\theta}\,, \\
    \beta_k(\theta) &=& \cos{\theta}\,\sin{k\theta}\,-\,k\,\sin{\theta}\,\cos{k\theta}\,, \\
    \hat{\beta}_k(\theta) &=& \sin{\theta}\,\sin{k\theta}\,+\,k\,\cos{\theta}\,\cos{k\theta}\,.
\end{eqnarray}

Another set of constraints comes from the flux ratios. Relative to
image $n$, we have
\begin{equation}
    f_{nl} = \frac{\mu_n}{\mu_l}.
\end{equation}
We then use (\ref{eqn:general G}), (\ref{eqn:inverse magnification})
and (\ref{eqn:expanded F}) to obtain
\begin{equation}
    \label{eqn:fluxratio}
    (f_{nl} - 1) r_n r_l = \frac{a_0}{2} \gamma_0(\theta_l) +
    \sum_{k=1}^{k_{\rm max}}\left[a_k\gamma_k(\theta_l) +
    b_k\delta_k(\theta_l)\right],
\end{equation}
where $l = 1, 2, \, \ldots \,, n - 1$. The functions $\gamma_k$ and
$\delta_k$ are defined by \citet{Evans_flux_ratios}, equation (18):
\begin{eqnarray}
    \gamma_k(\theta_l) &=& (1 - k^2)[f_{nl}r_l\cos{k\theta_n} - r_n\cos{k\theta_l}]\\
    \delta_k(\theta_l) &=& (1 - k^2)[f_{nl}r_l\sin{k\theta_n} - r_n\sin{k\theta_l}]\,.
\end{eqnarray}

We can combine equations (\ref{eqn:u equation}), (\ref{eqn:v
equation}) and (\ref{eqn:fluxratio}) into a single matrix equation:
\begin{equation}
    \label{eqn:matrix equation}
    \AMat \cdot \chiVec = \bVec,
\end{equation}
where $\chiVec$ is the ($2 k_{\rm max} + 1$)-dimensional vector of
parameters for which we are solving; $\chiVec = (u, v, a_0, a_2,
b_2, \ldots, a_{k_{\rm max}}, b_{k_{\rm max}})$. We drop $a_1$ and
$b_1$, which represent an unobservable translation of coordinates in
the source plane. The ($3n - 1$)-dimensional vector $\bVec$ contains
the observed image positions and flux ratios;
\begin{eqnarray}
\bVec &=& (x_1, \ldots, x_n, y_1, \ldots, y_n, \nonumber\\
&& (f_{n1}-1)r_n r_1, \ldots, (f_{n, n-1}-1)r_n r_{n-1}),
\end{eqnarray}
where $x_l = r_l \cos{\theta_l}$ and $y_l = r_l \sin{\theta_l}$. The
$(3n-1) \times (2k_{\rm max} + 1)$ matrix, $\AMat$, is defined in
equation (22) of \citet{Evans_flux_ratios}:
\begin{equation}
    \AMat = \left[\begin{array}{cccccccc}
    1 & 0 & \alpha_{01} & \alpha_{21} & \beta_{21} & \ldots \\
    \vdots & \vdots & \vdots & \vdots & \vdots & \vdots \\
    1 & 0 & \alpha_{0n} & \alpha_{2n} & \beta_{2n} & \ldots \\
    0 & 1 & \hat{\alpha}_{01} & \hat{\alpha}_{21} & \hat{\beta}_{21} & \ldots \\
    \vdots & \vdots & \vdots & \vdots & \vdots & \vdots \\
    0 & 1 & \hat{\alpha}_{0n} & \hat{\alpha}_{2n} & \hat{\beta}_{2n} & \ldots \\
    0 & 0 & \gamma_{01} & \gamma_{21} & \delta_{21} & \ldots \\
    \vdots & \vdots & \vdots & \vdots & \vdots & \vdots \\
    0 & 0 & \gamma_{0,n-1} & \gamma_{2,n-1} & \delta_{2,n-1} & \ldots \\
    \end{array}\right],
\end{equation}
where $\alpha_{kl}\equiv\alpha_k{(\theta_l)}$ etc.
\citet{Evans_flux_ratios} choose $k_{\rm max}$ such that $\AMat$ is
square. We then have $k_{\rm max} = 5$ for a $4$-image system
($n=4$). With this choice of $k_{\rm max}$ we can simply multiply
equation (\ref{eqn:matrix equation}) by $\AMat^{-1}$ to solve for
$\chiVec$, provided that $\AMat$ is non-singular.  To ensure
numerical stability, however, \citet{Evans_flux_ratios} solve for
$\chiVec$ using {\it singular value decomposition} (SVD).

\subsection{The Minimum Wiggle Model}
\label{sec:wiggle}

There are two main limitations of the method of
\citet{Evans_flux_ratios}.  On a technical point, their requirement
that $\AMat$ be square prevents one from probing the contributions
of higher-order multipoles.  More significantly, the effects of
external shear have not been included for several of the systems
they analyze. We now set out to address these two concerns.

In the case of arbitrary $\kmax$, SVD produces a particular solution
$\chiVec^{(0)}$ as well as a basis for the null space of $\AMat$: $
\{ \vect{\nu^{(i)}} \}$.  We then have a family of solutions,
\begin{equation}
\chiVec = \chiVec^{(0)} + \sum_{i = 1}^{N_p - N_c} c_i \,
\vect{\nu^{(i)}}, \label{eqn:paramList}
\end{equation}
where $N_p=(3n-1) > N_c=(2\kmax+1)$ are the numbers of parameters
and constraints, respectively.  We must now select appropriate
coefficients $c_i$ in order to construct the most plausible
solution. Since the lens galaxies we are considering are elliptical,
it seems reasonable to find the model with the smallest deviation
from elliptical symmetry.  In other words, we want to minimize the
wiggles in the isodensity contours.

For a curve of constant $\kappa$, the deviation $\delta r(\theta)$
from perfect elliptical symmetry is given by:
\begin{eqnarray}
\delta r(\theta) &\equiv& r(\theta) - r_0(\theta), \nonumber\\
&=& \frac{1}{2\kappa} \sum_{k = 3}^{k_{\rm max}} \left(1-k^2\right)
\left( a_k \cos k\theta + b_k \sin k\theta \right),
\label{eqn:deltar}
\end{eqnarray}
where
\begin{eqnarray}
r_0(\theta)=\frac{1}{2\kappa}\left[\frac{a_0}{2}-3( a_2 \cos 2\theta
+ b_2 \sin 2\theta)\right]
\nonumber
\end{eqnarray}
is an isodensity curve for a perfectly elliptical galaxy.  To
quantify the wiggles, we average $\delta r^2$ over $\theta$:
\begin{eqnarray}
\delta r^2_{\rm rms}\equiv \langle \delta r^2
\rangle_{\theta}=\frac{1}{8\kappa^2} \sum_{k = 3}^{k_{\rm max}}
(1-k^2)^2(a_k^2+b_k^2).
\end{eqnarray}
We are interested in the solution for which the root mean square
wiggle is minimized.

If we consider higher order multipoles but ignore shear, we simply
need to minimize the RMS wiggle with respect to the coefficients
$c_i$. Since $\langle \delta r^2 \rangle$ is quadratic in $a_k$ and
$b_k$, and hence also in $c_i$, this minimization is
straightforward.

When shear is included, the task of minimizing $\langle \delta r^2
\rangle$ becomes slightly more involved.  In particular, the lens
potential of equation (\ref{eqn:kappa and psi}) must be modified:
\begin{eqnarray}
\psi(r,\theta) = r F(\theta) - \frac{\gamma_1}{2} r^2 \cos{2 \theta}
- \frac{\gamma_2}{2} r^2 \sin{2 \theta},
\end{eqnarray}
for constants $\gamma_1, \gamma_2$. This modification requires that
the functions $\gamma_k$ and $\delta_k$ of equation
(\ref{eqn:fluxratio}), and the vector $\bVec$ of equation
(\ref{eqn:matrix equation}) be redefined by the expressions of
Appendix D of \citet{Evans_flux_ratios}. Namely,
\begin{eqnarray}
  \gamma_k(\theta_l) &=& (1 - k^2) \Bigl[ f_{nl}r_l\cos{k\theta_n} W(\theta_n)
    \nonumber\\
  &&\qquad\qquad - r_n\cos{k\theta_l} W(\theta_l) \Bigr] \\
  \delta_k(\theta_l) &=& (1 - k^2) \Bigl[ f_{nl}r_l\sin{k\theta_n} W(\theta_n)
    \nonumber\\
  &&\qquad\qquad - r_n\sin{k\theta_l} W(\theta_l) \Bigr]\,
\end{eqnarray}
where
\begin{equation}
    W(\theta_l) = 1+\gamma_1 \cos{2\theta_l}+\gamma_2 \sin{2\theta_l}
\end{equation}
and
\begin{eqnarray}
\bVec &=& [(1+\gamma_1)x_1+\gamma_2 y_1, \ldots, (1+\gamma_1)x_n+\gamma_2 y_n,
\nonumber \\
&& (1-\gamma_1)y_1+\gamma_2 x_1, \ldots, (1-\gamma_1)y_n+\gamma_2 x_n,
\nonumber \\
&& (f_{n1}-1)r_n r_1 (1-\gamma_1^2-\gamma_2^2), \ldots, \nonumber\\
&& (f_{n, n-1}-1)r_n r_{n-1} (1-\gamma_1^2-\gamma_2^2)].
\end{eqnarray}
We see that including $(\gamma_1,\gamma_2)$ as parameters to be
determined leads to a set of non-linear equations. To deal with this
problem, we use a non-linear optimization procedure. For specific
values of $(\gamma_1,\gamma_2)$, we can use SVD along with the
minimum wiggle criterion to solve for the source position and
Fourier coefficients. A minimization function can be employed to
find the optimal values for $(\gamma_1,\gamma_2)$.  We refer to the
resulting solution as the {\it minimum wiggle model}.

\section{Results}
\label{sec:results}

\begin{figure}
\begin{center}
\includegraphics[width=0.5\textwidth]{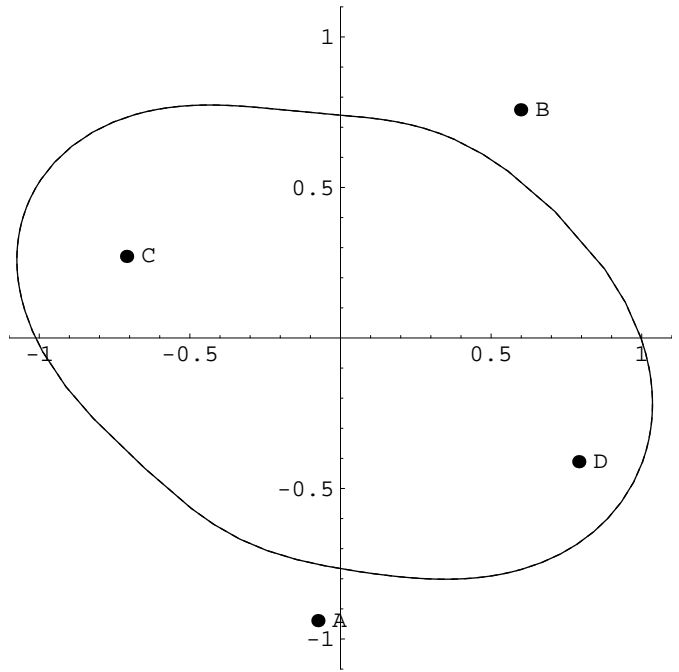}
\end{center}
\caption{Isodensity contour with $\kappa = 1/2$ (coincident with
critical curve) for Q2237+0305 with $\kmax=5$ and no shear. The axes
are labeled in arcseconds. \label{fig:2237}}
\end{figure}

Let us now apply our methods to five quadruply imaged quasars. We
begin with the Einstein cross, Q2237+0305, which does not exhibit
anomalous flux ratios at radio wavelengths, thus providing a simple
test case for the multipole expansion approach. Figure
\ref{fig:2237} shows a model with $\kmax = 5$ and no external shear
\citep[cf.][Figure 2]{Evans_flux_ratios}. The model exactly fits the
observational data presented by \citet{Falco_2237} and the CASTLES
website. The elliptical appearance of the isodensity contour
confirms that the multipole method can find reasonable solutions in
lens systems that do not require small-scale structure. Now let us
turn to four anomalous systems: B1422+231, B2045+265, B0712+472, and
B1933+503.

\subsection{External Shear}

\begin{figure*}
\begin{center}
\subfigure[Zero shear]
{\includegraphics[width=0.45\textwidth]{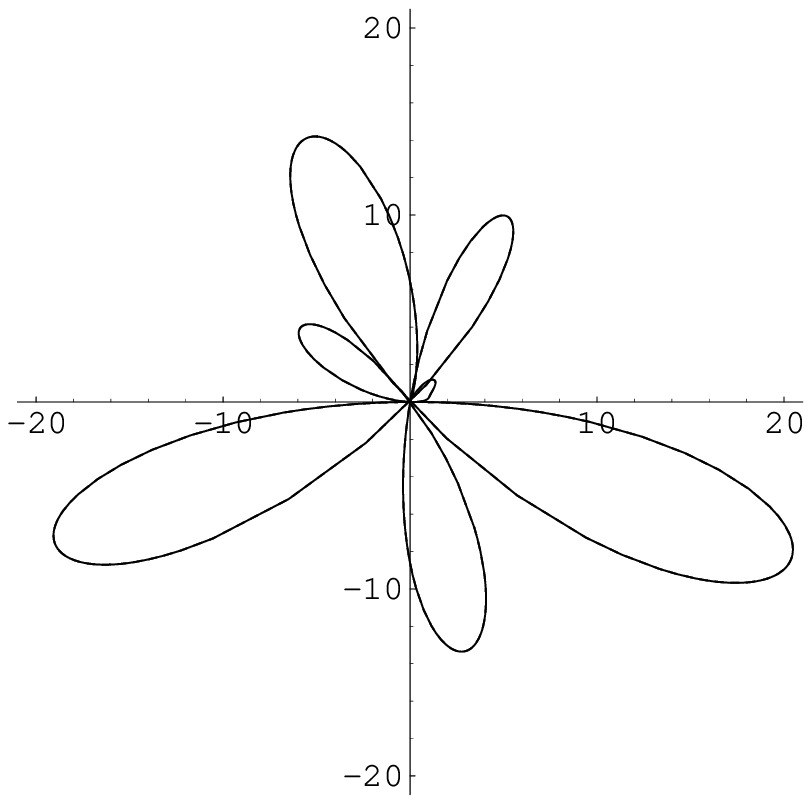} \label{fig: Zero
shear}} \hspace{0.1 in} \subfigure[Nonzero shear]
{\includegraphics[width=0.45\textwidth]{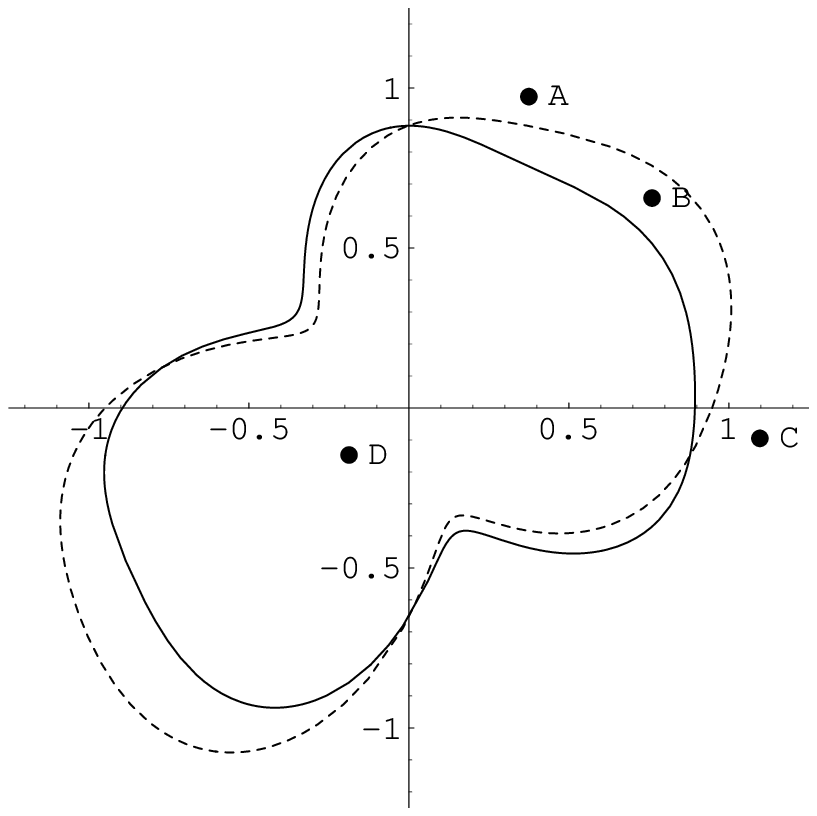} \label{fig:
Nonzero shear}}
\end{center}
\caption{Isodensity contours with $\kappa=1/2$ (solid) and critical
curves (dashed) for B1422+231. Panel (a) shows a model with
$\kmax = 5$ and no shear.  Panel (b) shows the solution for the
same value of $\kmax$, but nonzero shear parameters
$( \gamma_1 , \gamma_2) = (0.029, 0.170)$.  The dots show the image
positions (suppressed in panel (a) for clarity). Note that with
nonzero shear, isodensity contours and critical curves are not
identical. \label{fig:shearnoshear}}
\end{figure*}

\begin{figure*}
\begin{center}
\subfigure[$\kmax=9$]
{\includegraphics[width=0.45\textwidth]{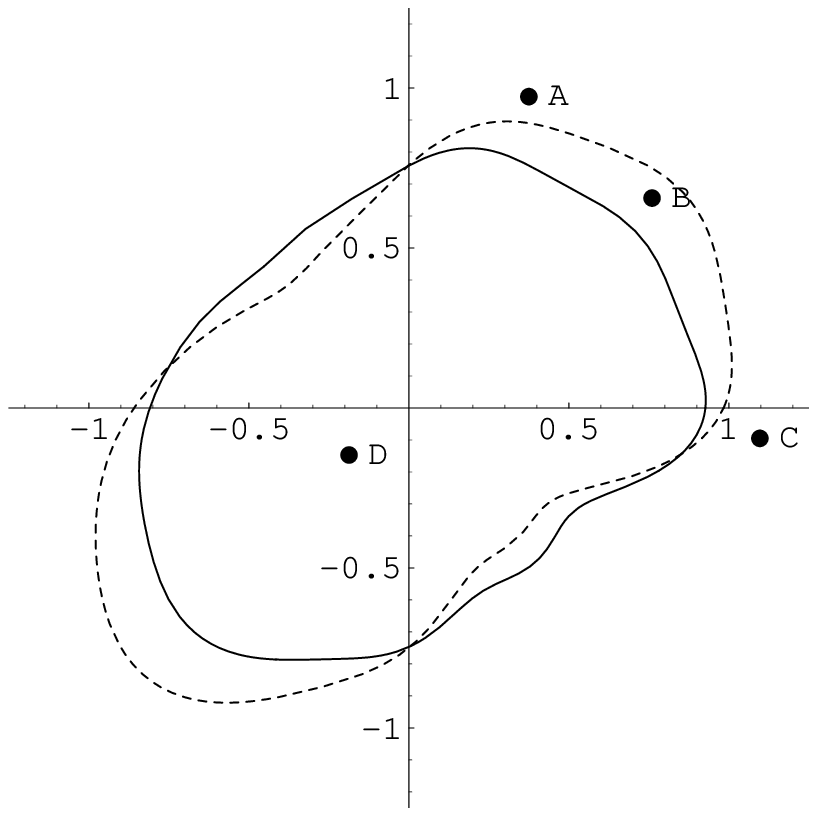}
\label{fig:kmax=9}} \hspace{0.1 in} \subfigure[$\kmax=25$]
{\includegraphics[width=0.45\textwidth]{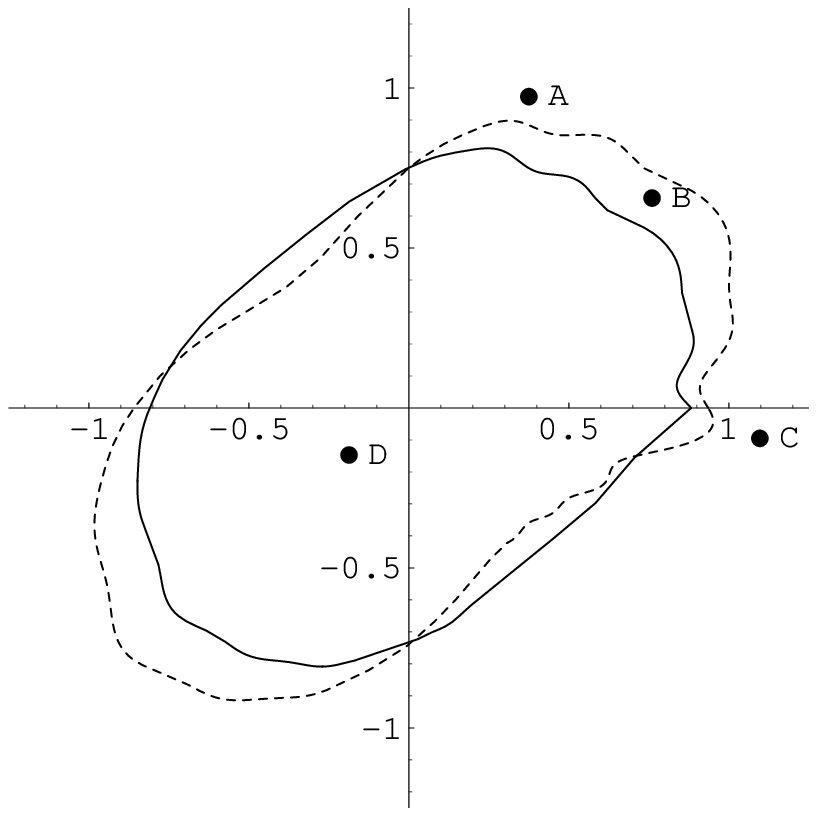}
\label{fig:kmax=25}}
\end{center}
\caption{Isodensity contours (solid) and critical curves (dashed)
for minimum wiggle models of B1422+231. Panel (a) shows a model for
$\kmax=9$ and shear parameters ($\gamma_1, \gamma_2$) = (0.028,
0.175). Panel (b) shows a model for $\kmax=25$ and shear parameters
($\gamma_1, \gamma_2$) = (0.030, 0.167). \label{fig:1422withshear}}
\end{figure*}

To motivate the need for external shear, we first study B1422+231
using a model that does not include shear.  To facilitate comparison
with the results of \citet{Evans_flux_ratios}, we consider a fifth
order multipole model, which fits the data of \citet{Impey_1422}
and \citet{Patnaik_1422} exactly. It is clear from
Figure \ref{fig: Zero shear} that this model is unphysical. In
addition to having a completely nonelliptical appearance, the model
isodensity contour crosses the origin indicating that $r$ becomes
negative.  \citet{Evans_flux_ratios} found a slightly better solution
by inflating the measurement uncertainties (see their Figure 6).
Even so, the model is incompatible with the observed structure of
elliptical galaxies, a point noted by the authors.

When we extend the model by including shear, it becomes possible to
find a physical --- albeit not necessarily plausible --- angular
structure for the lens galaxy (see Figure \ref{fig: Nonzero shear}).
Before we can draw any conclusions from this result, however, we
must determine whether the shear parameters and Fourier coefficients
we obtain are compatible with other observations. The shear
parameters we find for B1422+231 are reasonable, because the lens
lies in a group of galaxies that create a strong tidal field
\citep[see][]{Kundic_1422, Momcheva_tidal}. In particular, our shear
amplitude of $\gamma \equiv \sqrt{\gamma_1^2 + \gamma_2^2}=0.172$,
and orientation $\theta_\gamma \equiv(1/2)\tan^{-1}{(\gamma_2 /
\gamma_1)}=40^\circ$ are similar to those quoted by
\citet{Kundic_1422}.

The Fourier coefficients through order 4, plus some other model
properties, are given in Table \ref{tab:data}.  To interpret them,
we can determine the dimensionless octopole amplitude, $A_4$, which
describes the boxiness or diskiness of the isodensity contours, and
compare it with the octopole amplitudes measured for the isophotes
of elliptical galaxies.  The comparison is not perfect because lens
models involve the mass while observations involve the light, but we
can at least get a sense of whether the lens models are reasonable.
The octopole amplitude $A_4$ is just the Fourier coefficient for the
density, expressed in a coordinate frame aligned with the major axis
of the galaxy, and normalized by the semi-major axis length.  In
terms of the coefficients in Table \ref{tab:data}, the major axis
lies along the angle $\theta_2 = (1/2) \tan^{-1}(b_2/a_2)$. Rotating
into this coordinate frame then yields
\begin{equation}
  A_4 = -\frac{15}{R_{\rm ein}} \left( a_4 \cos 4\theta_2
    + b_4 \sin 4\theta_2 \right) .
\end{equation}
(The factor of $-15$ comes from $1-k^2$, which appears when we
convert from Fourier coefficients in the potential to those in the
density.)  If $A_4$ is negative (positive), the isodensity contours
are boxy (disky).  For B1422+231, our model with $\kmax = 5$ has
$A_4 = -0.056$.  For comparison, the octopole amplitudes for the
isophotes of real elliptical galaxies are in the range
$-0.015 \la A_4 \la 0.045$ \citep{Bender_a4,Saglia_a4}.  In other
words, the $\kmax = 5$ model is much more boxy than real galaxies
(which is not surprising in light of Figure \ref{fig: Nonzero shear}).
If we increase $\kmax$ to 9 (see below), we find $A_4 = -0.021$ which
is still rather boxy.  Going to $\kmax = 25$ yields $A_4 = -0.010$,
which is no more boxy than many observed elliptical galaxies.

\begin{table*}
\begin{center}
\caption{Fourier coefficients, normalized RMS wiggle, and shear
parameters for the various multipole lens models discussed in this
paper.  We quote models that fit the data exactly.  \label{tab:data}}
{\scriptsize
\begin{tabular}{c|lll|ll|lll|l}
System \vline & \multicolumn{3}{|c|}{B1422+231} \vline &
\multicolumn{2}{|c|}{B2045+265} \vline & \multicolumn{3}{|c|}{B0712+472} \vline & B1933+503 \\
\hline $\kmax$ & \multicolumn{1}{|c}{5} & \multicolumn{1}{c}{9} &
\multicolumn{1}{c|}{25} & \multicolumn{1}{c}{9} &
\multicolumn{1}{c|}{17} & \multicolumn{1}{c}{5} &
\multicolumn{1}{c}{6}
& \multicolumn{1}{c|}{7} & \multicolumn{1}{|c|}{8} \\
\hline\hline
$R_{\rm Ein}$ & \ 0.797 & \ 0.781 & \ 0.779 & \ 1.11 & \ 1.11 & \ 0.715 & \ 0.710 & \ 0.706 & \ 0.515 \\
$a_2$ & -0.0213 & -0.0159 & -0.0136 & \ 0.00158 & -0.0211 & -0.0306 & -0.0424 & -0.0361 & -0.0112 \\
$b_2$ & -0.0689 & -0.0492 & -0.0517 & \ 0.0139 & -0.0135 & \ 0.0394 & \ 0.0246 & \ 0.0239 & -0.0586 \\
$a_3$ & -0.00817 & -0.00359 & -0.00223 & -0.00928 & -0.000950 & -0.00333 & \ 0.000379 & -0.0000154 & \ 0.00144 \\
$b_3$ & \ 0.0132 & \ 0.000983 & \ 0.0000806 & -0.0159 & -0.00401 & -0.00595 & -0.000511 & \ 0.0000860 & -0.00143 \\
$a_4$ & -0.00226 & -0.00119 & -0.000685 & \ 0.000673 & \ 0.000580 & -0.00466 & -0.000709 & -0.000234 & \ 0.000869 \\
$b_4$ & \ 0.00198 & \ 0.000195 & -0.000196 & \ 0.00367 & -0.000128 & -0.000546 & \ 0.000113 & \ 0.000287 & -0.000358 \\ \hline $\delta r_{\rm rms}/R_{\rm Ein}$ & \ 0.166 & \ 0.0528 & \ 0.0376 & \ 0.235 & \ 0.0943 & \ 0.215 & \ 0.0993 & \ 0.0546 & \ 0.126 \\
\hline
$\gamma_1$ & \ 0.029 & \ 0.028 & \ 0.030 & \ 0.119 & \ 0.092 & -0.044 & -0.072 & -0.066 & -0.033 \\
$\gamma_2$ & \ 0.170 & \ 0.175 & \ 0.167 & \ 0.125 & \ 0.090 & -0.078 & -0.092 & -0.085 & \ 0.024
\end{tabular}
}
\end{center}
\end{table*}

\begin{figure*}
\begin{center}
\subfigure[$\kmax = 9$]
{\includegraphics[width=0.45\textwidth]{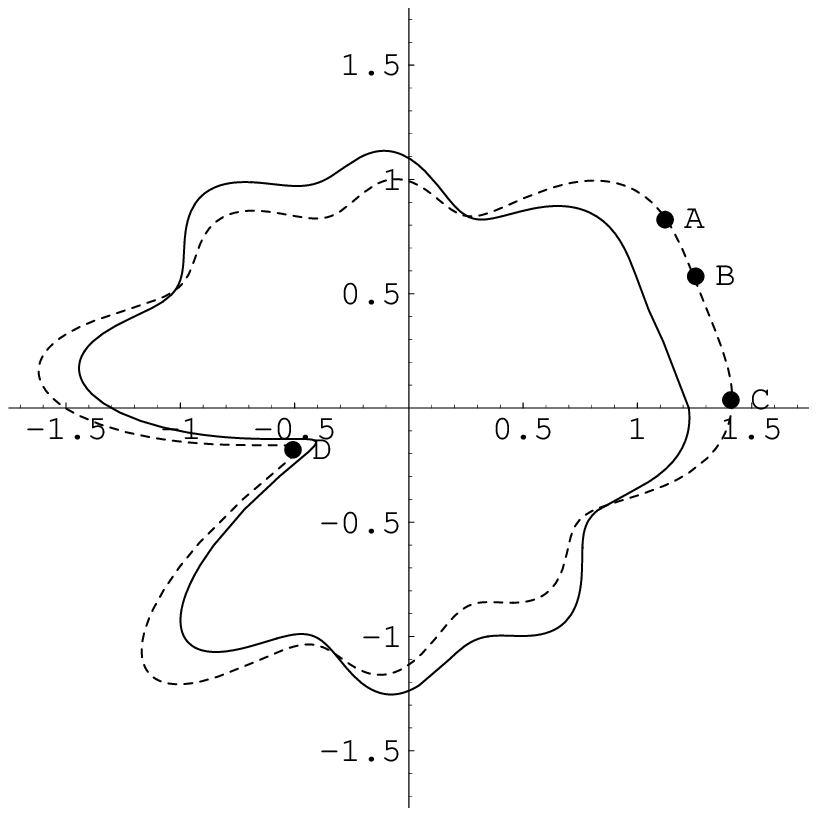}
\label{fig:2045_order9}} \hspace{0.1 in}
\subfigure[$\kmax = 17$]
{\includegraphics[width=0.45\textwidth]{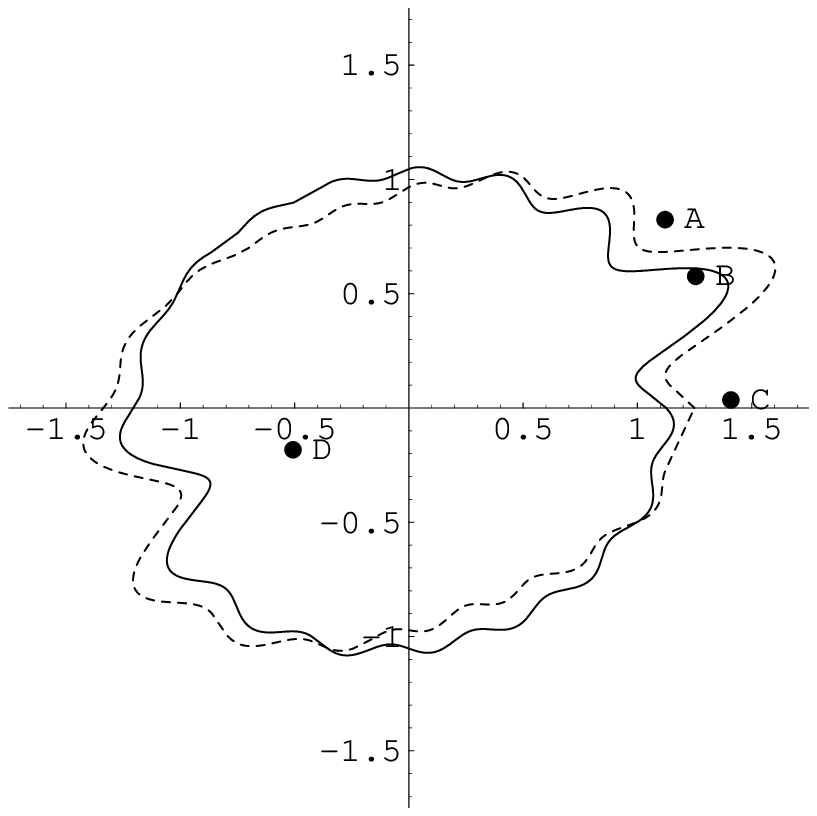}
\label{fig:2045_order17}}
\end{center}
\caption{Isodensity contours (solid) and critical curves (dashed)
for minimum wiggle models of B2045+265. Panel (a) shows the solution
with $\kmax = 9$ and ($\gamma_1, \gamma_2$) = (0.119, 0.125). Panel
(b) shows the solution with $\kmax = 17$ and ($\gamma_1, \gamma_2$)
= (0.092,0.090).\label{fig:2045}}
\end{figure*}

\subsection{Higher Order Multipoles}

Our next step is to include higher order multipole terms. In the
case of B1422+231, the lowest order series for which a somewhat
elliptical isodensity contour can be obtained is for $\kmax=9$ (see
Figure \ref{fig:kmax=9}). As we increase $\kmax$ the long-wavelength
components of galactic structure disappear in favor of wiggles that
are localized near the positions of the bright images A, B,
and C (see Figure \ref{fig:kmax=25}). In other words, away from the
images the model is smooth thanks to the minimum wiggle criterion.
But there must be small-scale structure in the vicinity of the
images in order to explain the observed flux ratios. Since the
wiggles in the isodensity contour are not dramatic, it is not clear
whether they should be interpreted as real features or just as
approximations of other sorts of structure (such as CDM clumps).

Let us now turn our attention to B2045+265. We find models that fit
the data of \citet{Fassnacht_2045} exactly. Unlike the case of
B1422+231, the isodensity contour we obtain for $\kmax=9$ is
completely unreasonable (see Figure \ref{fig:2045_order9}) and we
must include multipoles of order 17 to obtain a remotely plausible
model (see Figure \ref{fig:2045_order17}). Similar to B1422+231, we
find that including higher order multipoles reduces the RMS wiggle,
but pronounced deviations from ellipticity remain primarily near the
bright images. This suggests that the structure required to fit the
anomaly in B2045+265 truly is local to the images.

While the most obvious anomaly in B2045+265 is in the A/B/C triplet,
it is worth noting that our models also have small-scale structure
in the vicinity of the faint image D.  Dobler \& Keeton (2005) also
concluded that the flux of image D is puzzling, and suggested that
it has more to do with complex structure in the lens galaxy (such as
an isophote twist) than with substructure per se.  We cannot examine
that hypothesis here because our models are intrinsically
self-similar, but it will be interesting to keep this image in mind
as we develop more general models in the future.

\begin{figure}
\begin{center}
\includegraphics[width=0.45\textwidth]{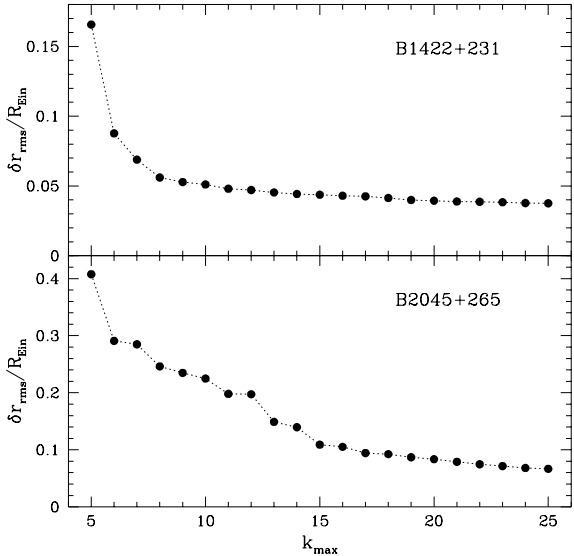}
\end{center}
\caption{Normalized RMS wiggle ($\delta r_{\rm rms}/R_{\rm Ein}$) as
a function of $\kmax$. \label{fig:wiggleplots}}
\end{figure}

When examining isodensity contours, it may not be completely obvious
that high-order models with small-scale undulations really have a
smaller total wiggle than low-order models.  To understand that,
recall that our minimum wiggle criterion (see eq.~21) is designed to
select the model whose isodensity contours deviate least from
elliptical symmetry.  As $k_{\rm max}$ increases, large-scale
departures from ellipticity can be traded for smaller-scale features
localized near the images in a way that does in fact decrease the
total wiggle.  Indeed, Figure \ref{fig:wiggleplots} shows that
$\delta r_{\rm rms}/R_{\rm Ein}$ does decrease monotonically as the
order of the multipole expansion increases.  The decrease is rapid
at small $k_{\rm max}$ but slows down as $k_{\rm max}$ increases,
and that gives us a sense of the order at which the multipoles have
basically converged to have the minimum amount of small-scale
structure.

\subsection{Measurement Uncertainties}

So far we have considered models that fit the data exactly, since
that is the natural outcome of an SVD analysis of the
underconstrained matrix equation (\ref{eqn:matrix equation}).
However, it is important to consider whether measurement
uncertainties affect our conclusions. While we could do this
analysis for B2045+265, we believe that the puzzling flux of image D
would complicate the interpretation. We turn instead to B0712+472.

We assume the image positions to be precise to within $\pm 3$ mas,
which is slightly conservative compared to the 1 mas uncertainties
claimed by \citet{Jackson_0712_position}. For the flux ratios, we
use the data of \citet{Koopmans_0712_flux}, who found the
uncertainties in the flux ratios of images B, C, and D relative to A
to be 7.2\%, 8.9\%, and 43\%, respectively. To be conservative, we
construct models for which we take the B/A and C/A uncertainties to
be 10\% and 20\% (we always use the observed flux uncertainty of
43\% for image D).

Since our formalism always produces models that fit the data
exactly, the way we include measurement uncertainties is to add
noise to the data and repeat our analysis.  For every run, we
perturb each data value by a random number drawn from a normal
distribution with the appropriate dispersion, and then fit our
model.  We repeat this process 100 times, and select the case that
has the smallest mean square wiggle. In this way we find the minimum
wiggle model that fits the data within the measurement
uncertainties.

\begin{figure*}
\begin{center}
\subfigure[$\kmax = 5$]
{\includegraphics[width=0.45\textwidth]{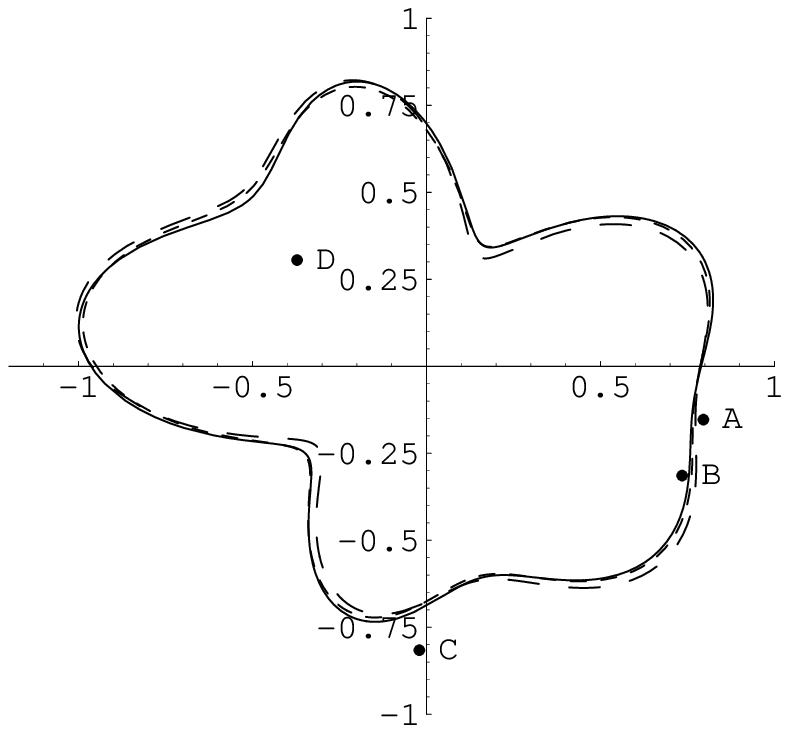}} \hspace{0.1 in}
\subfigure[$\kmax = 6$]
{\includegraphics[width=0.45\textwidth]{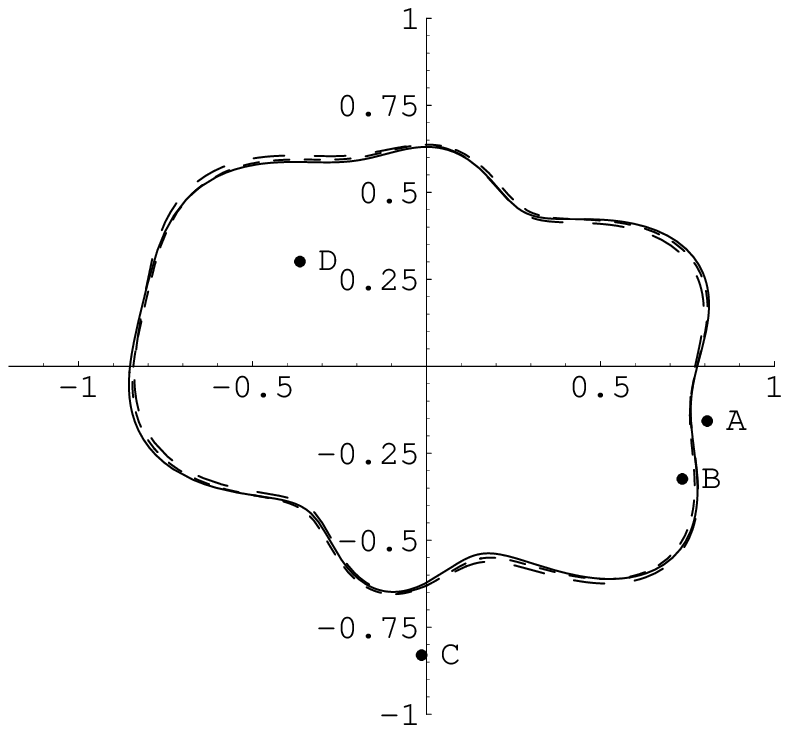}}
\subfigure[$\kmax = 7$]
{\includegraphics[width=0.45\textwidth]{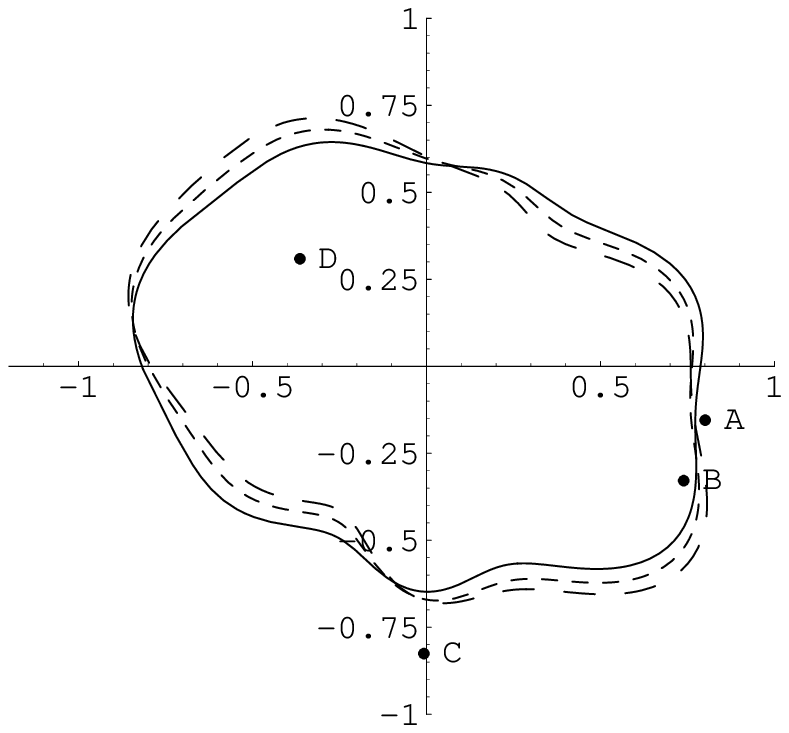}}
\end{center}
\caption{Model isodensity contours for B0712+472 with increasing
multipole order. The solid curve in each panel fits the data
exactly. The short and long-dashed curves fit the data to within
10\% and 20\% flux uncertainties, respectively (except for faint
image D, whose flux uncertainty is fixed at the observed value of
43\%; see \citealt{Koopmans_0712_flux}).
\label{fig:uncertainty}}
\end{figure*}

Figure \ref{fig:uncertainty} shows the results for B0712+472. As we
would expect, including measurement uncertainties produces models
with slightly smaller wiggles.  However, the changes are not
significant enough to transform an implausible model into an
acceptable solution.

\subsection{A Multi-Source Lens Model}

To conclude this section, we study B1933+503, where two sources are
lensed by a single galaxy into two four-image configurations
\citep[][and references accompanying their Table 1]{Cohn_1933}. Both
the image positions and flux ratios corresponding to one source are
known, while only the image positions of the second source have been
determined to reasonable precision. We therefore have $N_c =
2(n_{1}+n_{2})+n_{1}-1 = 19$ constraints, compared with $N_c =
3n-1=11$ for the other systems we have analyzed, where
$n_{1}=n_{2}=4$. Since we must now fit a second source position, our
list of parameters increases by two: $N_p = 2\kmax + 3$. The matrix
$\AMat$ then has the dimensions
$[2(n_1+n_2)+n_1-1]\times(2\kmax+3)$, and is given by adding more
rows that represent the additional position constraints, as follows:
\newline

\begin{equation}
    \AMat = \left[\begin{array}{cccccccc}
    1 & 0 & 0 & 0 & \alpha_{01}^{(1)} & \alpha_{21}^{(1)} & \beta_{21}^{(1)} & \ldots \\
    \vdots & \vdots & \vdots & \vdots & \vdots & \vdots & \vdots & \vdots \\
    1 & 0 & 0 & 0 & \alpha_{0 n_1}^{(1)} & \alpha_{2 n_1}^{(1)} & \beta_{2 n_1}^{(1)} & \ldots \\
    \\
    0 & 1 & 0 & 0 & \hat{\alpha}_{0 1}^{(1)} & \hat{\alpha}_{2 1}^{(1)} & \hat{\beta}_{2 1}^{(1)} & \ldots \\
    \vdots & \vdots & \vdots & \vdots & \vdots & \vdots & \vdots & \vdots \\
    0 & 0 & 1 & 0 & \alpha_{0 1}^{(2)} & \alpha_{2 1}^{(2)} & \beta_{2 1}^{(2)} & \ldots \\
    \vdots & \vdots & \vdots & \vdots & \vdots & \vdots & \vdots & \vdots \\
    0 & 0 & 1 & 0 & \alpha_{0 n_2}^{(2)} & \alpha_{2 n_2}^{(2)} & \beta_{2 n_2}^{(2)} & \ldots \\
    \\
    0 & 0 & 0 & 1 & \hat{\alpha}_{0 1}^{(2)} & \hat{\alpha}_{2 1}^{(2)} & \hat{\beta}_{2 1}^{(2)} & \ldots \\
    \vdots & \vdots & \vdots & \vdots & \vdots & \vdots & \vdots & \vdots \\
    0 & 0 & 0 & 0 & \gamma_{0 1}^{(1)} & \gamma_{2 1}^{(1)} & \delta_{2 1}^{(1)} & \ldots \\
    \vdots & \vdots & \vdots & \vdots & \vdots & \vdots & \vdots & \vdots \\
    0 & 0 & 0 & 0 & \gamma_{0,n_1 - 1}^{(1)} & \gamma_{2,n_1 - 1}^{(1)} & \delta_{2,n_1 - 1}^{(1)} & \ldots \\
    \end{array}\right]
\end{equation}

We first consider just the primary set of four images, which are
labeled 1, 3, 4, and 6.  (There is a flux ratio anomaly such that
image 4 is brighter than expected.)  The minimum wiggle model with
$k_{\rm max}=8$ that fits these images is not very plausible --- it
has a large protrusion near image 4, and a smaller one near image 1
--- but at least the density is positive definite.  When we add the
additional constraints from the positions of images 2a, 2b, 5, and
7, they dramatically reduce the solution space ($k_{\rm max} = 8$ is
the lowest order case that is not over-constrained).  The
only models that remain are unphysical.  Modestly increasing $k_{\rm
max}$ does not help.  In other words, multipole models cannot
simultaneously fit the anomalous fluxes of images 1, 3, 4 and 6, and
the positions of images 2a, 2b, 5 and 7.

\begin{figure*}
\begin{center}
\subfigure[Including image positions and flux ratios for images 1,
3, 4, and 6 only] {\includegraphics[width=0.45\textwidth]{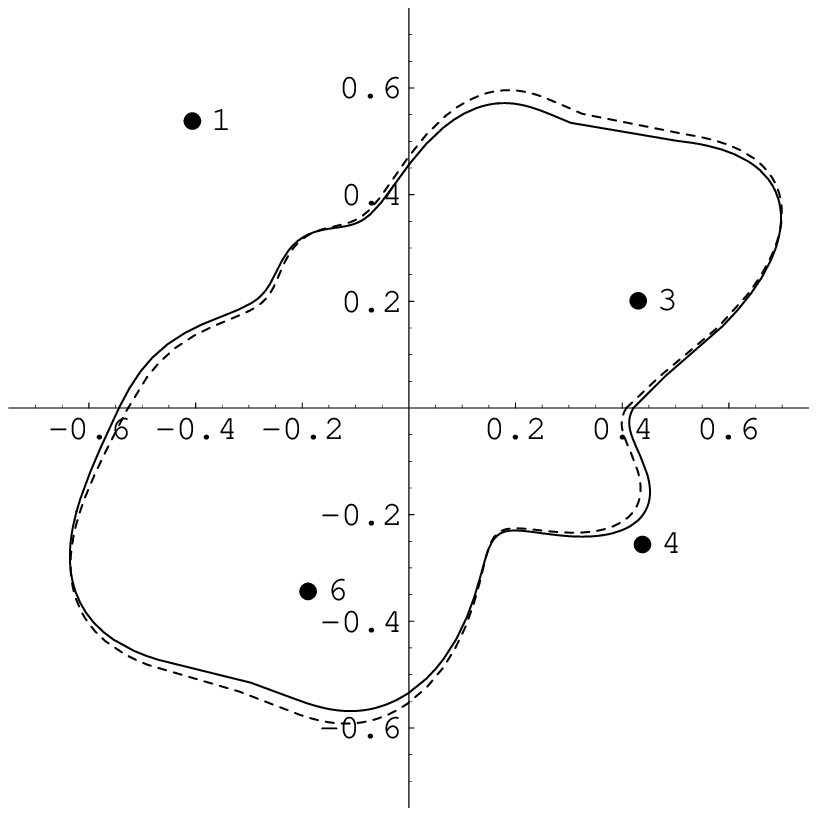}
\label{fig:1933_1346}} \subfigure[Including all constraints]
{\includegraphics[width=0.45\textwidth]{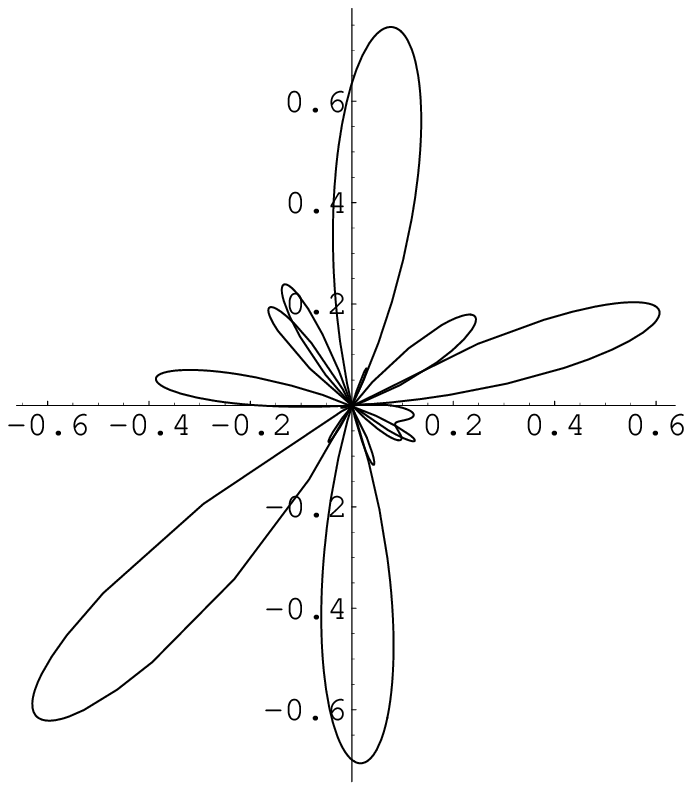}
\label{fig:1933_all}}
\end{center}
\caption{Isodensity contours (solid) and critical curves (dashed)
for two models of B1933+503. Panel (a) includes the constraints for
images 1, 3, 4 and 6. Panel (b)  shows a model in which all of the
observational constraints have been fit. Both models account for
external shear and multipoles up to $\kmax=8$. \label{fig:1933}}
\end{figure*}

\section{Conclusions}

We have shown that extending the method of \citet{Evans_flux_ratios}
by including external shear and higher-order Fourier terms is
essential for understanding whether multipole models can fit
observed lenses. The results for B1422+231 we have obtained are of a
very different character from those of \citet{Evans_flux_ratios}.
Even so, we cannot conclude that the multipole approach provides an
acceptable explanation of flux ratio anomalies.

The system B2045+265 requires multipoles of order $\ga 15$. Even
this level of small-scale structure leads to a rather wiggly angular
dependence of density near the three bright images. Next, our
analysis of B1933+503 reveals that a fundamental difficulty exists
in fitting a multi-source lens with a simple multipole model.
Finally, our method naturally finds models that fit the data
exactly, but we have shown that our conclusions are not very
sensitive to measurement uncertainties.

Our results suggest that there is a more fundamental problem with
the global approach taken by \citet{Evans_flux_ratios} and ourselves
in the current paper. It is possible that the problem simply comes
from our choice of small scale structure. Sines and cosines provide
a useful but by no means unique basis for carrying out a series
expansion of the angular part of the potential. In addition, we have
assumed a particular form for the radial dependence --- notably
self-similarity --- which may need to be modified in order to find
an acceptable galactic density function. While the present analysis
seems to rule out simple multipole models, the question of whether
CDM clumps provide the only plausible solution has yet to be fully
answered.

\section*{Acknowledgements}

We thank Arthur Kosowsky, Tad Pryor, and Jerry Sellwood for
helpful discussions. ABC is supported by an NSF Graduate Research
Fellowship.

\end{document}